\documentclass[12pt]{article}
\usepackage{latexsym, amsmath, epsfig, amsfonts, amsbsy}
\DeclareFontFamily{OT1}{rsfs10}{}
\DeclareFontShape{OT1}{rsfs10}{m}{n}{ <-> rsfs10 }{}
\DeclareMathAlphabet{\mathscript}{OT1}{rsfs10}{m}{n}

\if@twoside\oddsidemargin25mm
\else\oddsidemargin25mm\evensidemargin25mm\marginparwidth25mm\fi
\def\Z{\mathbb{Z}}
\def\C{\mathbb{C}}

\def\R{\mathbb{R}}

\newcommand{\ft}[2]{{\textstyle\frac{#1}{#2}}}
\def\brr{\begin{equation}}
\def\err{\end{equation}}
\def\brr{\begin{eqnarray}}
\def\err{\end{eqnarray}}
\def\ba{\left(\begin{array}}
\def\ea{\end{array}\right)}
\def\lf{\left.\begin{array}{c}}
\def\rf{\end{array}\right.}

\newcommand{\dr}{\raise.3ex\hbox{$\stackrel{\leftarrow}{\partial }$}{}}
\newcommand{\dl}{\raise.3ex\hbox{$\stackrel{\rightarrow}{\partial}$}{}}
\newcommand{\topi}{\raise.3ex\hbox{$\stackrel{\pi}{\longrightarrow}$}{}}
\newcommand{\ns}{\normalsize}

\renewcommand{\a}{\alpha}
\renewcommand{\b}{\beta}

\newcommand{\g}{\gamma}

\begin{document}


\begin{titlepage}

\vspace{-4cm}

\title{
   \hfill{\ns CU-TP-1026\,,\,HU-EP-01/29\,,\,UPR-950T\\}
   \hfill{\ns hep-th/0108078\\[2cm]}
   {\LARGE Four-Dimensional $N=1$ Super Yang-Mills Theory \\
   from an {\it M}-Theory Orbifold \\[1cm]  }}

\author{{\bf
   Charles F. Doran$^{1}$ ,\,
   Michael Faux$^{1}$
   and Burt A.~Ovrut$^{2}$}\\[5mm]
   {\it $^1$Departments of Mathematics and Physics} \\
   {\it Columbia University} \\
   {\it 2990 Broadway, New York, NY 10027} \\[3mm]
   {\it $^2$Department of Physics, University of Pennsylvania} \\
   {\it Philadelphia, PA 19104--6396, USA}}
\date{}

\maketitle

\vspace{.3in}

\begin{abstract}
\noindent
Gravitational and gauge anomalies provide stringent constraints
on which subset of chiral models can effectively describe
{\it M}-theory at low energy. In this paper,
we explicitly construct an abelian orbifold of {\it M}-theory to obtain an
$N=1$, chiral super Yang-Mills theory in four dimensions, using
anomaly matching to determine the entire gauge and representation structure.
The model described in this paper is the simplest four dimensional model
which one can construct from {\it M}-theory compactified on an abelian
orbifold without freely-acting involutions.
The gauge group is
$SO(12)\times SU(8)\times SU(2)\times SU(2)\times U(1)$.

\vspace{.5in}
\noindent
\end{abstract}

\thispagestyle{empty}

\end{titlepage}


\section{Introduction}

Despite the fact that a fundamental description of {\it M}-theory
has so far remained elusive, it is nevertheless possible to describe
interesting and predictive aspects of its effective phenomenology.
This is possible because,
whatever {\it M}-theory turns out to be, it should relate
at low energy to eleven-dimensional supergravity.  This
statement is actually quite powerful, especially
when eleven-dimensional spacetime has a topology
involving a compact
orbifold factor, owing to rigorous constraints derivable from
requirements imposed by field theory
gauge and gravitational anomaly cancellation.

There has been quite a lot of recent interest in $M$-theory on
$G_2$ holonomy seven-manifolds for the construction of $N=1$ supersymmetric
theories in four dimensions \cite{Acharya,AtWitt}.
In this paper we describe, in microscopic detail, a particular
$N=1$ model associated with a particularly interesting {\it M}-theory orbifold
\footnote{Some authors refer to our quoteint space as an ``orientifold" because
the finite group action includes a parity flip.  We prefer a more broad use of the
term ``orbifold", since ``orientifold" has
a slightly different connotation in string theory.}.
Although our model involves compactification on a seven-manifold, it has
the structure of a $T^6/(\Z_2 \times \Z_2)$ orbifold (admitting
a Calabi-Yau resolution, \cite{joyce}) times a closed interval
$S^1/\Z_2$.  Such
an ``orbifold with boundary" falls outside of the class of $G_2$-resolvable
orbifolds of $T^7$ studied by Joyce \cite{joyce}
\footnote{We refer particularly to Definition 6.5.1 and
all of Chapter 11 in Joyce's book \cite{joyce}.}.
Nevertheless, by compactification on this orbifold
we do obtain a four-dimensional, $N=1$, chiral super Yang-Mills theory
from eleven-dimensional supergravity via explicit cancellation of anomalies.

In our analysis, we impose strict anomaly cancelation at each point in the
eleven-dimensional spacetime.  This is a substantially more restrictive
criterion than that implied by anomaly cancellation on smaller
spaces obtained when compact dimensions shrink to zero size.  The latter
circumstance accesses only what we call the collective anomaly, whereas
our approach involves a more ``microscopic" picture of the
localized states.  Admittedly, in our approach we are able to compute only
those chiral states needed to cancel anomalies.  Any additional
localized states which can be added without introducing additional
anomalies are invisible to our analytic probe. Typically, this
redundancy is of very limited scope, however.  In this paper we
describe a consistent microscopic description of the localized
states in one particular orbifold.  Furthermore, as explained in
\cite{phase}, one expects a hierachy of consistent solutions
for the gauge group and representation content associated with a
given orbifold compactification of {\it M}-theory.  Typically,
sets of such consistent solutions are linked by phase transitions,
often mediated by fivebranes and small instanton transformations.
In this paper we discuss only one particular solution; we fully expect that
this solution describes only one corner of a more robust and interesting
moduli space.  Having said this, we remark that the particular
{\it M}-theory orbifold analyzed in this paper had been previously mentioned in
\cite{gopmuk} as an interesting N=1 model, where it appeared as Model (C',C').
However, in that paper no attempt was made to describe the associated spectrum
from a microscopic point of view.  The fact that the gauge group described
in that paper differs from ours is not particulaly troubling either, since
we are describing a different corner of moduli space.  It would be an
interesting exercise to provide a microscopic description of the
models described in \cite{gopmuk}.  To the best of our knowledge this
current paper is the only extant microscopic description of a
$D=4$ $N=1$ model derived from {\it M}-theory.

The reason why anomaly cancellation is important in the context
of {\it M}-theory orbifolds is that the the lift of the
action of the orbifold quotient group to the gravitino field
generically serves to project that field chirally onto
even-dimensional fixed-point loci.  On fixed-planes of
dimensionality ten or six this projection induces
gravitational anomalies, owing to the chiral coupling of the
gravitino
to currents which are classically conserved due to
classical reparameterization invariance of the fixed-planes.
However, since all local anomalies
must cancel, the presence of gravitino-induced fixed-plane
anomalies allows one to infer additional structure,  such as
Yang-Mills supermultiplets living on the fixed-planes,
or specialized electric and magnetic sources of $G$ flux
\footnote{Using standard notation,
$G$ is the four-form field strength living in the bulk
supergravity multiplet.}, since these supply necessary
contributions to the overall anomaly, either quantum mechanically
or as ``inflow", so as to render the theory consistent.

In generic situations
orbifold fixed-point loci can be complicated, involving
fixed-planes of various dimensionalities which can intersect.
As a result there are additional concerns owing to
gauge anomalies and mixed anomalies induced by chiral projection
of the gaugino fields in the Yang-Mills supermultiplets
onto fixed-plane intersections.  Because of this, the
cancellation of all anomalies typically involves an elaborate
conspiracy of quantum contributions and inflow contributions.
In a previous series of papers \cite{mlo, phase, hetk3, chern},
much of the technology needed for implementing
anomaly matching in orbifolds with intersecting fixed-planes was developed.
Complementary aspects and several physical observations about
these orbifolds were also addresed by other authors, \cite{ksty} and
\cite{bds} in particular.
This work extended the ideas and technology implemented in simpler
orbifolds involving only isolated (i.e. non-intersecting) fixed-planes
described by
Ho{\v r}ava and Witten \cite{hw1,hw2} and by Dasgupta and Mukhi \cite{dasmuk}.
Until now, the only orbifolds with intersecting fixed-planes which have been
analyzed are those corresponding to topologies
$\R^6\times S^1/\Z_2\times K3$ in which the $K3$ factor degenerates
to a global orbifold $K3\to T^4/\Z_M$, for the four possible cases
$M=2,3,4$ and $6$.  The effective theory in these previously studied cases,
obtained in the limit that the radii of the compact dimensions
becomes very small, are uniformly six-dimensional.

In this paper we describe a new example of an {\it M}-theory orbifold
which has a four-dimensional effective description, and also has four
dimensional fixed-planes.  This model exhibits a pretty feature
in that each four-dimensional fixed-plane lies at the mutual intersection
of three orthogonal six-dimensional fixed-planes, each of which is a
sub-manifold of a ten-dimensional fixed-plane.  What is nice about this
feature is that the constraints imposed by ten- and six-dimensional
gravitational anomaly cancellation
impinge directly on the structure of the effective
field theory in four dimensions, despite the fact that there
are no gravitational anomalies specifically in four
dimensions.  This is possible because the four-planes in question are very
special sub-manifolds
of the fixed ten-planes and also of the fixed six-planes.
This introduces gravitational anomalies into four dimensional physics
in a novel manner.  Thus, despite the fact that the model which we
present is not of immediate phenomenological interest,
it does indroduce a powerful formalism for deriving
four-dimensional physics from {\it M}-theory.

In our model, eleven-dimensional
spacetime has topology $\R^4\times S^1/\Z_2\times T^6/(\Z_2)^2$.
This orbifold is of the Ho{\v r}ava-Witten variety, since it includes
$S^1/\Z_2$ as a factor, and has
quotient group $(\Z_2)^3$. Since
the orbifold is of the Ho{\v r}ava-Witten variety, in order to cancel
the ten-dimensional anomaly, the fixed ten-planes,
of which there are two, each support $E_8$ Yang-Mills supermultiplets.
In this case, however, there are additional
gravitational anomalies induced on the six-planes.   In order to cancel
these, it is necessary to introduce hypermultiplets living on these
six-planes.
Furthermore, it is necessary that the ten-dimensional $E_8$ gauge
group is broken to a subgroup $E_8\to {\cal G}_\a\subset E_8$ on
the six-dimensional fixed-plane corresponding to the element $\a$ of $(\Z_2)^3$.
This symmetry breakdown is codified by the action of the quotient group
on the $E_8$ root lattice and is, in effect, a
description of a small $E_8$ gauge instanton which
is localized on the fixed-plane.

In our model, the cancellation of anomalies
on the ten-dimensional fixed-planes
and also on the six-dimensional fixed-planes proceeds exactly
as described in \cite{hetk3} for the case
of the $S^1/\Z_2\times T^4/\Z_2$ compactification.
This is because these planes are locally identical to the
analogs in that simpler construction.
The novel feature of the $(\Z_2)^3$ model, however, derives from the fact that
the six-planes intersect each other at four-planes.
As a result of this, there are important consistency requirements
which control the ultimate breakdown patterns of $E_8$ as one
approaches first a six-plane and then moves along that six plane
and lands on the four plane intersection.
In this paper, we present an explicit consistency analysis
which enables us to compute the complete set of twisted states and the
gauge group localized on the four-dimensional fixed-planes.  We do not
review the Ho{\v r}ava-Witten analysis \cite{hw2} which explains the ten-dimensional
anomaly cancellation nor do we review the cancellation of the
six-dimensional anomalies.  The interested reader is referred to
\cite{mlo, phase, hetk3, chern} for a comprehensive description
of these cases.

Section 2 addresses the global geometric aspects of our model, describing the
explicit action of $(\Z_2)^3$ which gives rise to our orbifold of $T^7$.  In the
next section, the analysis of the local anomaly for the $(Z_2)^2$ orbifold in
\cite{phase, hetk3} is extended to this case.  Although the story for the
ten-planes is quite analogous, the six-planes require more subtle methods.
For this reason we introduce ``branching tables'' and ``embedding diagrams''
for determining which projections in the $E_8$ root lattice are compatible
with the orbifold quotient group action.  In Sections 4 and 5 we use this data
to determine the spectrum seen by the four-dimensional intersection which arises
from the ten-dimensional $E_8$ fields, and discuss further twisted states
which we need to introduce to cancel the six-dimensional anomalies.  Finally,
following a synopsis, we indicate why the four-dimensional gauge anomaly does
not arise for our orbifold, and summarize the representation content of the
chiral multiplets in our model (``Model 1'') in Table 7.  Further models with
$D=4$ and $N=1$ SUSY, based on both abelian and non-abelian orbifolds, will be
discussed in forthcoming work.

\section{Global Geometric Aspects}

Consider a spacetime with topology $\R^4\times T^7$,
where the compact $T^7$ factor is described by
the quotient $\R^7/\Lambda$ with lattice
$\Lambda$.
Parameterize the compact dimensions
using three complex coordinates $\{z_1,z_2,z_3\}$ and one real coordinate
$x^{11}$.
The orbifold which defines our model is obtained from this
torus by making additional
identifications described by parity
flips as shown in Table \ref{modone}.  The sole condition on the lattice
$\Lambda$ in $\C^3 \oplus \R$ is that these parity flips induce lattice
automorphisms.
A minus sign in that table implies a relative overall sign change
of the indicated coordinate (the column header) by the
indicated element (the row header).
Each row in
Table \ref{modone} describes a generator of the full quotient
group, which is $(\Z_2)^3$.
 \begin{table}
 \begin{center}
 \begin{tabular}{c|cccc}
 & \hspace{.3in} & \hspace{.3in} & \hspace{.3in} &  \\[-.3in]
 & $z_1$ & $z_2$ & $z_3$ & $x^{11}$ \\[.1in]
 \hline
 $\a$     & $-$ & $+$ & $-$ &  $-$ \\[.1in]
 $\b$     & $-$ & $-$ & $+$ &  $-$ \\[.1in]
 $\g$     & $+$ & $-$ & $-$ &  $-$ \\[.1in]
 \end{tabular} \\[.2in]
 \caption{The action of the quotient group $(\Z_2)^3$ on the
 compact coordinates $\{z_1,z_2,z_3,x^{11}\}$
 corresponding to the model described in the text.}
 \label{modone}
 \end{center}
 \end{table}

The quotient group $(\Z_2)^3$ has order eight, with elements
$\{1,\a,\b,\g,\a\b,\b\g,\a\g,\a\b\g\}$.  The
corresponding fixed-planes have real dimensionalities
$\{11,6,6,6,7,7,7,10\}$, respectively,  where we have
included the four non-compact dimensions in this
accounting. The $\a\b\g$-invariant ten-planes have multiplicity
two, and correspond to hypersurfaces $x^{11}=0$ and $x^{11}=\pi$.
We refer to these ten-planes as $M^{10}_1$ and $M^{10}_2$, respectively.

\begin{figure}
 \begin{center}
 \includegraphics[width=3.5in,angle=0]{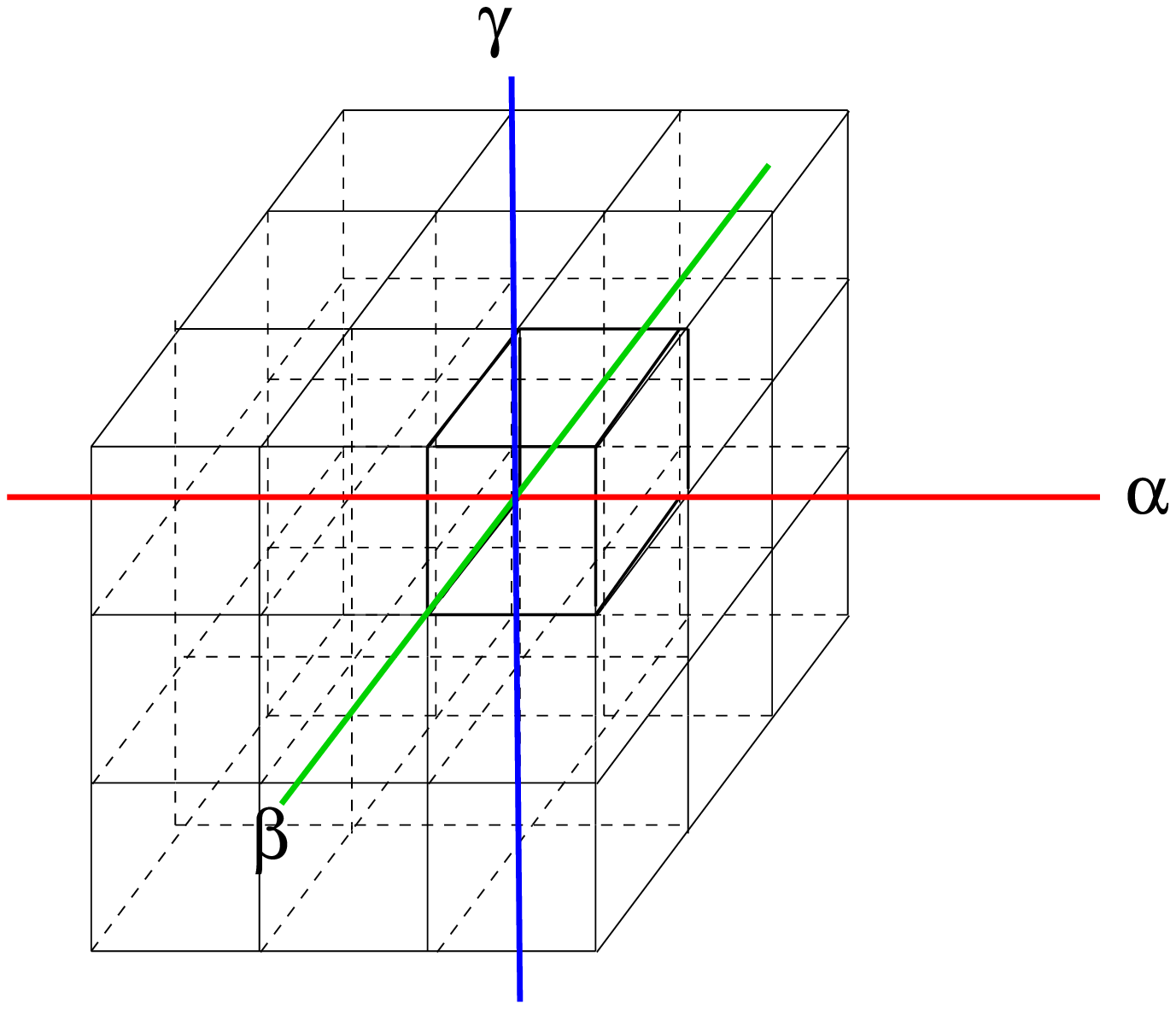}\\[.2in]
 \caption{A depiction of the orbifold described in the text with the eleventh
 dimension suppressed.  This picture is entirely within one of the
 two $\a\b\g$ ten-planes, and illustrates the sixteen $\a$
 six-planes, the sixteen $\b$ six-planes and the sixteen $\g$
 six-planes within that $\a\b\g$ ten-plane.}
 \label{boxbox}
 \vspace{1in}
 \includegraphics[width=3.5in,angle=0]{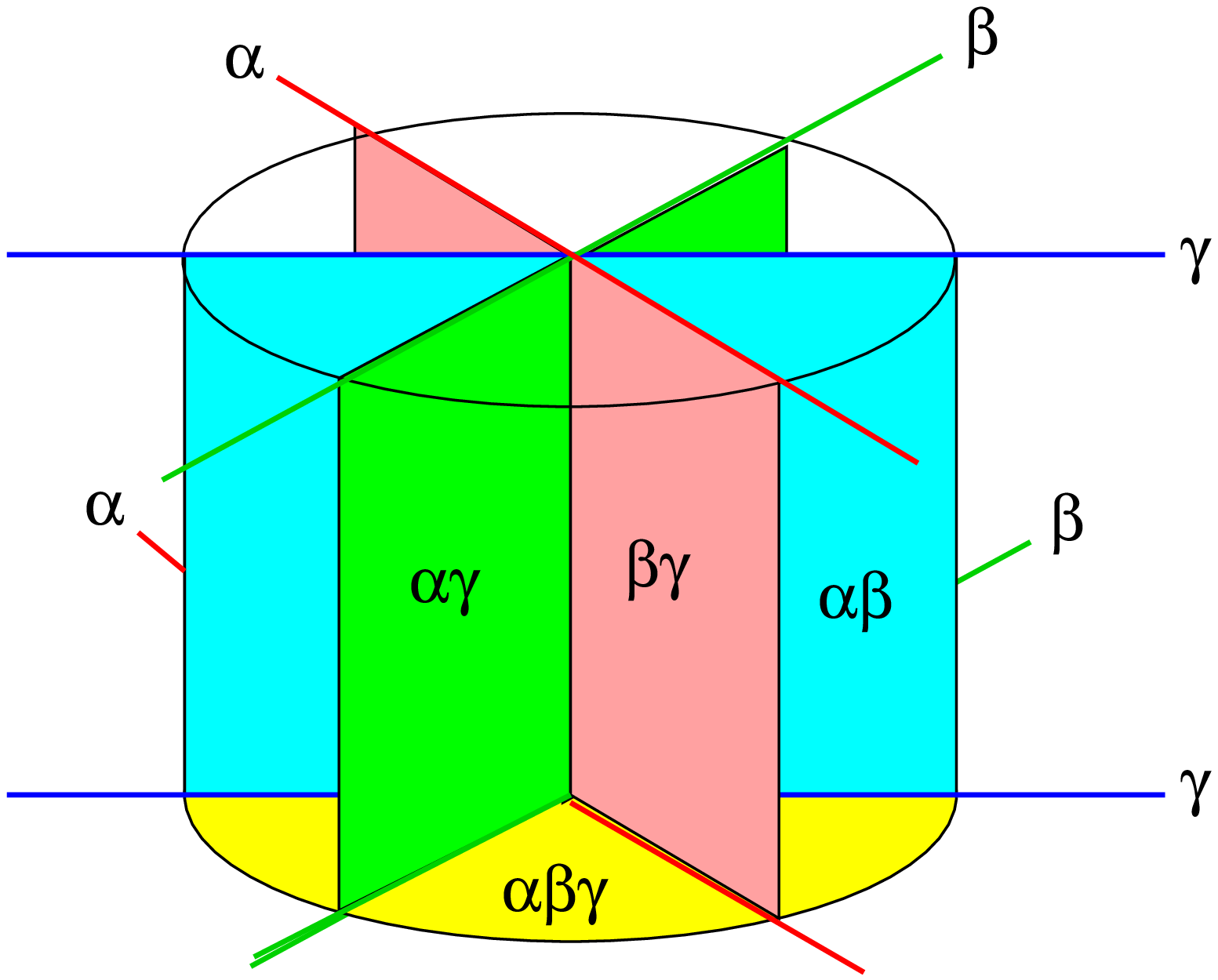}\\[.2in]
 \end{center}
 \caption{A local depiction showing one
 of the sixty-four five-planes, including the eleventh dimension,
 and showing three of the seven planes, one each of $\a\b$, $\b\g$
 and $\a\g$.}
 \label{waterwheel}
 \end{figure}

The $\a$-invariant six-planes, the $\b$-invariant six-planes
and the $\g$-invariant six-planes each have multiplicity 32, sixteen
of which are submanifolds of $M^{10}_1$  and
sixteen of which are sub-manifolds of $M^{10}_2$.
Within a given ten-plane $M^{10}_i$ there are 64 parallel
four-planes which are each invariant under $\a$, $\b$, and $\g$,
each describing a mutually transversal intersection of one $\a$-plane,
one $\b$-plane and
one $\g$-plane.   The global geometry of the six-planes within a given
ten-plane are conveniently
depicted in Figure \ref{boxbox}, where we have suppressed the
$x^{11}$ dependance of the non-compact
coordinates.

The $\a\b$-invariant seven-planes have multiplicity
sixteen.  Each of these interpolates between one $\g$-invariant
six-plane (a submanifold of $M^{10}_1$ ) and another $\g$-invariant six-plane
(a submanifold of $M^{10}_2$), with the interpolation parameterized
by $x^{11}$.  Similarly, sixteen $\b\g$-planes interpolate
between $\a$-planes and sixteen more $\a\g$-planes interpolate between
$\b$-planes.  The seven-planes triply intersect at 64 five-planes
which are each invariant under $\a\b$, $\b\g$, and $\a\g$.
Each five-plane interpolates between two of the
four planes described above.

The global geometry is conveniently displayed as
shown in Figure \ref{waterwheel}, which depicts the geometry near one
of the five-planes as it interpolates between two of the four-planes.
Figure \ref{waterwheel} includes representations of every sort of
fixed-plane, and every sort of intersection which occurs in this orbifold.
Since the collection of planes as drawn in Figure \ref{waterwheel} resemble
a sort of waterwheel, we refer to such a diagram as a waterwheel
diagram.  These figures are especially useful for maintaining perspective
during the ensuing analysis.

\section{Local Anomaly Cancellation}

As mentioned in the introduction, the geometry of the
ten-planes and of the six-planes for the orbifold described
in the previous section are locally identical to
analogous fixed-planes in the simpler $(\Z_2)^2$ orbifold
described in \cite{phase, hetk3}.  As a result, we can apply
certain results from the analyses described in those papers.
In particular, the ten-planes must each support a ten-dimensional $E_8$
Yang-Mills multiplet.   The situation
regarding the six-planes is more complicated, however.
We start by reviewing the situation in the simpler
$(\Z_2)^2$ orbifold, and then describe extra constraints which
pertain to the $(\Z_2)^3$ case.  For simplicity, in this paper we consider
only the possibility that there are no {\it M}-fivebranes in the bulk of
the orbifold.

For the case of the $S^1/\Z_2\times T^4/\Z_2$ orbifold
\cite{phase, hetk3},
we can locally cancel the six-dimensional anomalies in either
of two distinct ways. In the first case a given six-plane
has magnetic charge
$-1/4$, while lattice reflections
describe a breakdown $E_8\to E_7\times SU(2)$ as the six plane is approached.
In the second case, the six-plane has magnetic charge $+1/4$
and lattice reflections describe a different breakdown $E_8\to SO(16)$.
The ambiguity is resolved, however, by global constraints
derived by integrating the $dG$ Bianchi identity.  These require
that the $E_7\times SU(2)$ solution and the $SO(16)$ solutions
be paired, with one occurring on six-planes within $M^{10}_1$,
and the other occurring in
complementary six-planes inside $M^{10}_2$.

For the case of the $(\Z_2)^3$ orbifold, we have three
types of six-planes which respectively correspond to
the elements $\a$, $\b$ and $\g$ described above.  On a given six-plane
the gravitational anomaly can be cancelled locally via the same
two choices described in the previous paragraph.  But we need to
re-think the global constraints in this context, owing to the
relative complexity of the global fixed-plane network.
To be concrete, we focus on the neighborhood of one of the
four-dimensional intersection vertices, such as the one
distinguished in Figure \ref{boxbox} as the intersection
of the three highlighted six-planes.  This same four-plane
is also depicted as the upper point in Figure \ref{waterwheel} where all of the
depicted planes converge.   On each of the three
six-planes which mutually intersect at the given four-plane,
there are two possible choices for the local gauge
subgroup ${\cal G}_i$.  Here ${\cal G}_i$ corresponds
to ${\cal G}_\a$, ${\cal G}_\b$ or ${\cal G}_\g$ depending on
which six plane is being considered.
In each case,
the two possible choices are related to the two possible choices of
magnetic charges, since these correlate with the associated
$E_8$ lattice reflection, under which
$E_8\to {\cal G}_i\subset E_8$.

The three six-planes being considered are all submanifolds of a particular ten-plane
fixed under the triple product $\a\b\g$.  Thus, the three six-dimensional gauge groups
${\cal G}_\a, {\cal G}_\b$ and ${\cal G}_\g$ are each subgroups of
the same $E_8$.
These subgroups are separately fixed under the respective
actions of $\a, \b$,  and $\g$ on the $E_8$ root lattice.
 An important observation is
that the entire $E_8$ lattice must remain fixed under the triple product $\a\b\g$.
Otherwise the $E_8$ group would be broken
at generic points on the ten-manifold, which would irreparably spoil the
ten-dimensional anomaly.
As a result of this, given the action of $\a$ and $\b$ on the $E_8$
lattice, the action of $\g$ is fixed.  Specifically $\g$ must act on the
$E_8$ lattice precisely as the product $\b^{-1}\a^{-1}=\a\b$. (The equality
follows because $\a$ and $\b$ each generate $\Z_2$, and are therefore self-inverses,
and because the quotient group is abelian.) This
uniquely ensures that $\a\b\g$ acts trivially on the $E_8$ lattice.

\begin{figure}
 \begin{center}
 \includegraphics[width=2.5in,angle=0]{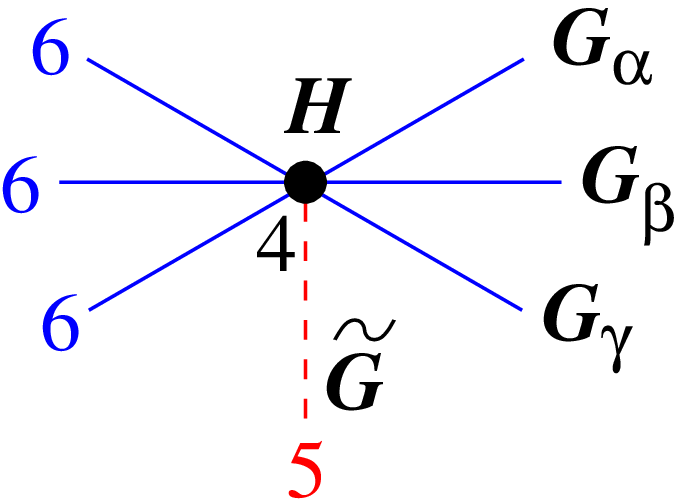}
 \caption{One of the four-dimensional intersection vertices}
 \label{int4}
 \vspace{1in}
 \includegraphics[width=2.5in,angle=0]{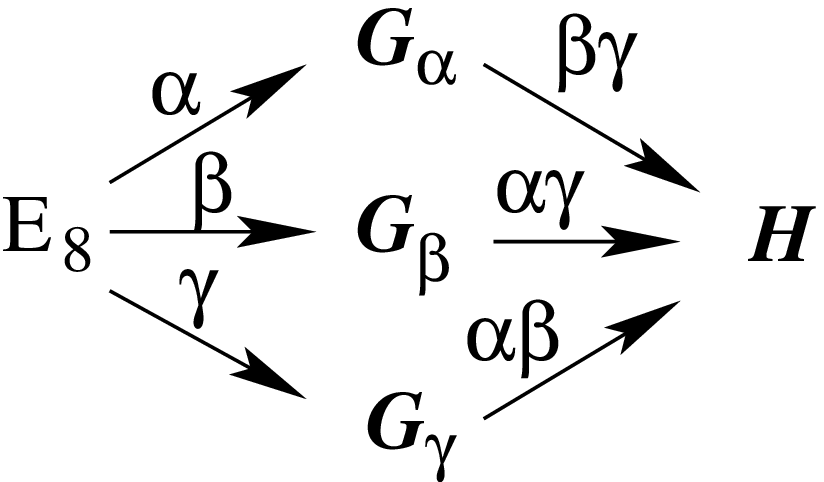}
 \end{center}
 \caption{The overlapping breakdown pattern $E_8\to {\cal H}$}
 \label{break}
 \end{figure}

Subject to the constraint described in the previous
paragraph, we need to determine how $\b$ and $\g$ act on ${\cal G}_\a$,
(i.e. how these elements are realized as reflections on the
sublattice of $E_8$ corresponding to the root lattice of
${\cal G}_\a$), and similarly how $\a$ and
$\g$ act on ${\cal G}_\b$ and how $\a$ and $\b$ act on
${\cal G}_\g$.    Since there is a unique subgroup
${\cal H}\subset E_8$ which remains
invariant under $\a$, $\b$, and $\g$, it follows that
the three groups ${\cal G}_\a$, ${\cal G}_\b$ and ${\cal G}_\g$
must each break down to the same group
${\cal G}_i\to {\cal H}$ under the lattice projections described above.
The state of affairs is illustrated by Figures
\ref{int4} and \ref{break}.

In Figure \ref{int4} we see a
depiction of the three six-planes mutually intersecting at
the four-plane under discussion.  The four-plane is illustrated by the
heavy dark spot.  This figure shows the physical
geometry of the fixed-plane intersection.  The gauge groups
corresponding to the various fixed-planes are also included in this
figure, as are the dimensionalities of the planes
\footnote{Note that we have
also drawn, using a dotted line, the five-plane described previously, and have
indicated that this, too, can potentially support its own Yang-Mills
multiplet, with gauge group $\tilde{G}$.}.
In Figure \ref{break} we see a depiction of the
group theoretic branchings from $E_8$ to the subgroup $H$, in which
the consistency of the actions of the six elements of $(\Z_2)^3$ other
than the identity and the triple product $\a\b\g$ is apparent.

We want to determine which sets of
three subgroups ${\cal G}_\a$, ${\cal G}_\b$ and ${\cal G}_\g$ of
$E_8$ can consistently overlap to satisfy the situation
illustrated in Figures \ref{int4} and \ref{break}.  To do this, we first
isolate the candidate subgroups of $E_8$ associated with each of
the six-planes $\a$, $\b$ and $\g$.
For the case at hand, the candidate subgroups are either
$E_7\times SU(2)$ or $SO(16)$.  Thus, in
this case, ${\cal G}_\a$, ${\cal G}_\b$ and ${\cal G}_\g$ are each
selected from between these two choices. But this must be done
subject to the constraint that $\a\b\g$ leaves the entire
$E_8$ group invariant.  This places a restriction on which {\it combinations} of
selections are permitted.

There is a systematics which resolves the
consistent breakdown pattern.
Given a pair of subgroups ${\cal G}_\a$ and ${\cal G}_\b$,
we compare maximal subgroups of these to see if we can find any
matches.  If there is a match, then this is a candidate for the group $H$.
If there aren't any matches, then we refine the
search by including subgroups at the next depth, i.e. include the
maximal subgroups of maximal subgroups.
For example: if we select ${\cal G}_\a=E_7\times SU_2$
and ${\cal G}_\b=SO_{16}$, then there is a unique depth-one
candidate for $H$, namely $SO_{12}\times SU_2\times SU_2$.

We define a preliminary breaking pattern as a set
$\{\,{\cal G}_\a\,,\,{\cal G}_\b\,;\, H\,\}$ meeting the above criteria.
The qualifier ``preliminary" reminds us that we have to verify the
consistency of such an ansatz, as we explain below.
Given a preliminary breaking pattern, we determine
the branching pattern for the ${\bf 248}$ representation
of $E_8$ according to $E_8\to {\cal G}_\a\to H$ and
$E_8\to {\cal G}_\b\to H$.  For illustration, we choose
$\{{\cal G}_\a\,,\,{\cal G}_\b\,|\,H\,\}=
\{E_7\times SU_2\,,\,SO_{16}\,|\,SO_{12}\times SU_2\times SU_2\,\}$.
In this case, the $\a$ plane involves the following branching
\brr \a: E_8 &\to& E_7\times SU_2
     \nonumber\\[.1in]
     &\to& SO_{12}\times SU_2\times SU_2
     \nonumber\\[.15in]
     {\bf 248} &\to& ({\bf 133},{\bf 1})\oplus
     ({\bf 1},{\bf 3})\oplus
     ({\bf 56},{\bf 2})
     \label{ba}\\[.1in]
     &\to& [\,({\bf 66},{\bf 1},{\bf 1})\oplus
     ({\bf 1},{\bf 3},{\bf 1})\oplus
     ({\bf 32},{\bf 2},{\bf 1})\oplus
     ({\bf 1},{\bf 1},{\bf 3})\,]\oplus
     ({\bf 32}',{\bf 1},{\bf 2})\oplus
     ({\bf 12},{\bf 2},{\bf 2}) \,.
     \nonumber
\err
In this tabulation we have obtained the branching rules from
\cite{slansky} and/or \cite{patera}.  (In the final line we have
kept square brackets around the terms in the decomposition
corresponding to the adjoint representation of
$E_7\times SU_2$, for reasons to become apparent.)
Similarly, the $\b$ plane  involves the following branching
\brr \b: E_8 &\to& SO_{16}
     \nonumber\\[.1in]
     &\to& SO_{12}\times SU_2\times SU_2
     \nonumber\\[.15in]
     {\bf 248} &\to& {\bf 120}\oplus {\bf 128}
     \label{bb}\\[.1in]
     &\to& [\,({\bf 66},{\bf 1},{\bf 1})\oplus
     ({\bf 1},{\bf 3},{\bf 1})\oplus
     ({\bf 1},{\bf 1},{\bf 3})\oplus
     ({\bf 12},{\bf 2},{\bf 2})\,]\oplus
     ({\bf 32},{\bf 2},{\bf 1})\oplus
     ({\bf 32}',{\bf 1},{\bf 2}) \,,
     \nonumber
\err
where we have employed the same systematics as described above for
the $\a$-plane branching.

We would like to express the group theoretic branching described above
in terms of the orbifold quotient group acting on the $E_8$ root lattice.
For the case at hand this is a relatively simple exercise, since all
of the elements of $(\Z_2)^3$ independently square to the identity.
As a consequence, we can realize the relevant group actions as
reflections on some subset of the $E_8$ root vectors.  For example,
to realize a branching $E_8\to E_7\times SU(2)$ it suffices to leave invariant
136 root vectors corresponding to roots of an $E_7\times SU(2)$ subgroup
of $E_8$, and to invert the remaining 112 root vectors.
Similarly to realize a branching $E_8\to SO(16)$ we leave invariant
120 root vectors corresponding to an $SO(16)$ subgroup and invert the
remaining 128.  Some of the consistency issues discussed above
translate into issues pertaining to the permissible consistent choices of
sublattices which can be acted upon by group elements $\a$, $\b$
and $\g$ in particular ways.

\subsection{Branching Tables}

A useful tool for collecting relevant
data in order to study consistent lattice projections is
a special table, which we will call a ``branching table".  In such a table
we partition the $E_8$ lattice vectors, row-wise, according to representations
of the subgroup $H$.  We then construct three columns, one corresponding
to the element $\a$, one to $\b$ and one to $\g$.  We fill out the
table by placing in each slot either a plus sign or a minus sign.
A plus sign indicates that the group element corresponding to the column
leaves invariant those root vectors corresponding to the row.  A minus
sign indicates that the indicated group element reflects the associated
vectors across the origin of the root space.

 \begin{table}
 \begin{center}
 \begin{tabular}{|c||c|c|c|}
 \hline
 & \hspace{.7in} & \hspace{.7in} & \hspace{.7in} \\[-.1in]
 {\bf 248} & $E_7\times SU_2$ & $SO_{16}$ & $E_7\times SU_2$ \\[.1in]
 \hline
 ({\bf 66},{\bf 1},{\bf 1})  &  + & + & + \\[.1in]
 ({\bf 1},{\bf 3},{\bf 1})   &  + & + & + \\[.1in]
 ({\bf 1},{\bf 1},{\bf 3})   &  + & + & + \\[.1in]
 ({\bf 12},{\bf 2},{\bf 2})  &  $-$ & $+$ & $-$ \\[.1in]
 ({\bf 32}',{\bf 1},{\bf 2}) &  $-$ & $-$ & + \\[.1in]
 ({\bf 32},{\bf 2},{\bf 1})  &  $+$ & $-$ & $-$ \\[.1in]
 \hline
 \end{tabular}\\[.3in]
 \caption{Branching table describing $E_8\to SO_{12}\times
 SU_2\times SU_2$.}
 \label{branch1}
 \vspace{.5in}
 \begin{tabular}{|c||c|c|c|}
 \hline
 & \hspace{.7in} & \hspace{.7in} & \hspace{.7in} \\[-.1in]
 {\bf 248} & $SO_{16}$ & $E_7\times SU_2$ & $SO_{16}$ \\[.1in]
 \hline
 ${\bf 63}_0$  &  + & + & + \\[.1in]
 ${\bf 1}_0$   &  + & + & + \\[.1in]
 ${\bf 70}_0$ & $-$ & $+$ & $-$ \\[.1in]
 ${\bf 1}_{+2}\oplus{\bf 1}_{-2}$   &  $-$ & $+$ & $-$ \\[.1in]
 ${\bf 28}_1\oplus{\bf\overline{28}}_{-1}$  &  $+$ & $-$ & $-$ \\[.1in]
 ${\bf 28}_{-1}\oplus{\bf\overline{28}}_{+1}$  &  $-$ & $-$ & $+$ \\[.1in]
 \hline
 \end{tabular}\\[.3in]
 \caption{Branching table describing $E_8\to SU_8\times U_1$.}
 \label{branch2}
 \end{center}
 \end{table}

Given the branching patterns described in (\ref{ba}) and
(\ref{bb}) we construct the branching table, according to the
above prescription, as shown in Table \ref{branch1}.
The entries in the first row of Table \ref{branch1} describes the
way that the generator $\a$ acts on the $E_8$ root vectors,
which are partitioned into representations of $H=SO(12)\times SU(2)\times SU(2)$.
There are a total of 136 invariant root vectors in the first row (i.e. those with
plus signs).  These describe the
root system of a particular $E_7\times SU(2)$ subgroup of $E_8$.
This can be verified from the branching rules describing
$E_7\to SO(12)\times SU(2)$.
The second column of Table \ref{branch1} corresponds to the element $\b$.
Statements analogous to those made about the first column verify that
the 120 invariant vectors listed in this column describe the
root system of a particular $SO(16)$ subgroup of $E_8$.

The third column of Table \ref{branch1} describes the element $\g$.
>From the discussion above, we know that this element does not act in
an independent manner on the $E_8$ lattice.
Instead, $\g$ acts in the same way as the product $\a\b$.
Thus, the third row of a given branching table, is obtained by
multiplying the first column with the second column. The fact that
$\{E_7\times SU(2),SO(16)\,|\,SO(12)\times SU(2)\times SU(2)\}$ is consistent
is then verified by the fact that the action of $\g$ which appears in
Table \ref{branch1} does, in fact, reconcile as the root system of another
$E_7\times SU(2)$ subgoup of $E_8$.   Consistency is ensured
because $E_7\times SU(2)$ is one of
the two consistent choices of $E_8$ subgroups.  (Consistent here means that
the associated six-dimensional anomalies can be cancelled.)

In this paper we are describing a relatively simple example
in which the group actions on the lattice are never more complicated
than mere reflections.  In future papers, however, we will describe more
interesting
scenarios in which the quotient group is realized relatively
nontrivially on the lattices.  The branching tables which we
are introducing in this paper have a natural generalization in those cases.

For the consistent triple branching illustrated
in Table \ref{branch1}, $\a$, $\b$ and $\g$ collectively involve two branchings
to $E_7\times SU(2)$ and one branching to $SO(16)$.  Owing to global
considerations discussed above, there must
exist a consistent complimentary scenario, involving two instances
of $SO(16)$ and one instance of $E_7\times SU(2)$.
Since the former
posibility is {\it defined} by the choice $H=SO(12)\times SU(2)\times SU(2)$,
there must exist an alternate choice for $H$.  Since
$SO(12)\times SU(2)\times SU(2)$ was the unique common depth-one subgroup
of both $E_7\times SU(2)$ and $SO(16)$, we need to go to greater depth
in order to find an alternate preliminary breaking pattern.
At the next depth there is again a unique choice, namely
$\{SO(16),E_7\times SU(2)\,|\,SU(8)\times U(1)\}$, so that
$H=SU(8)\times U(1)$.  This second case can be analyzed precisely as above,
with results summarized as in Table \ref{branch2}.
(Note that it is not possible to describe a consistent triple branching involving
three
instances of $E_7\times SU(2)$ or three instances of $SO(16)$, as can be
easily verified by trying to construct corresponding  branching
tables.)

We assume that branching decribed by Table \ref{branch1} corresponds to
a four-plane intersection inside of $M^{10}_1$, which we will
call the ``upstairs" region, and that the branching described by
Table \ref{branch2} corresponds to a four-plane intersection inside of
$M^{10}_2$, which we will call the ``downstairs" region.

\subsection{Embedding Diagrams}

\begin{figure}
 \begin{center}
 \vspace{.5in}
 \includegraphics[width=2.5in,angle=0]{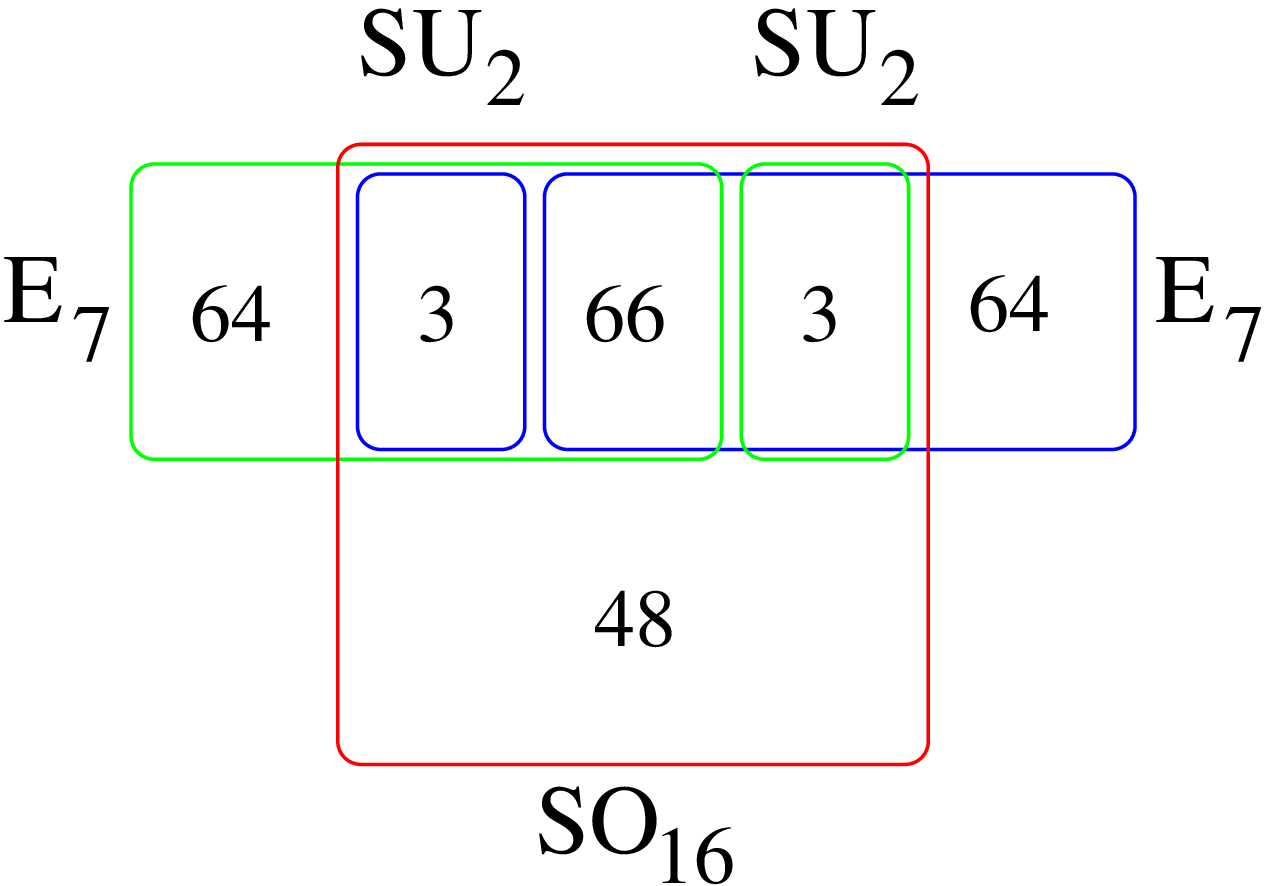} \hspace{.4in}
 \includegraphics[width=2.5in,angle=0]{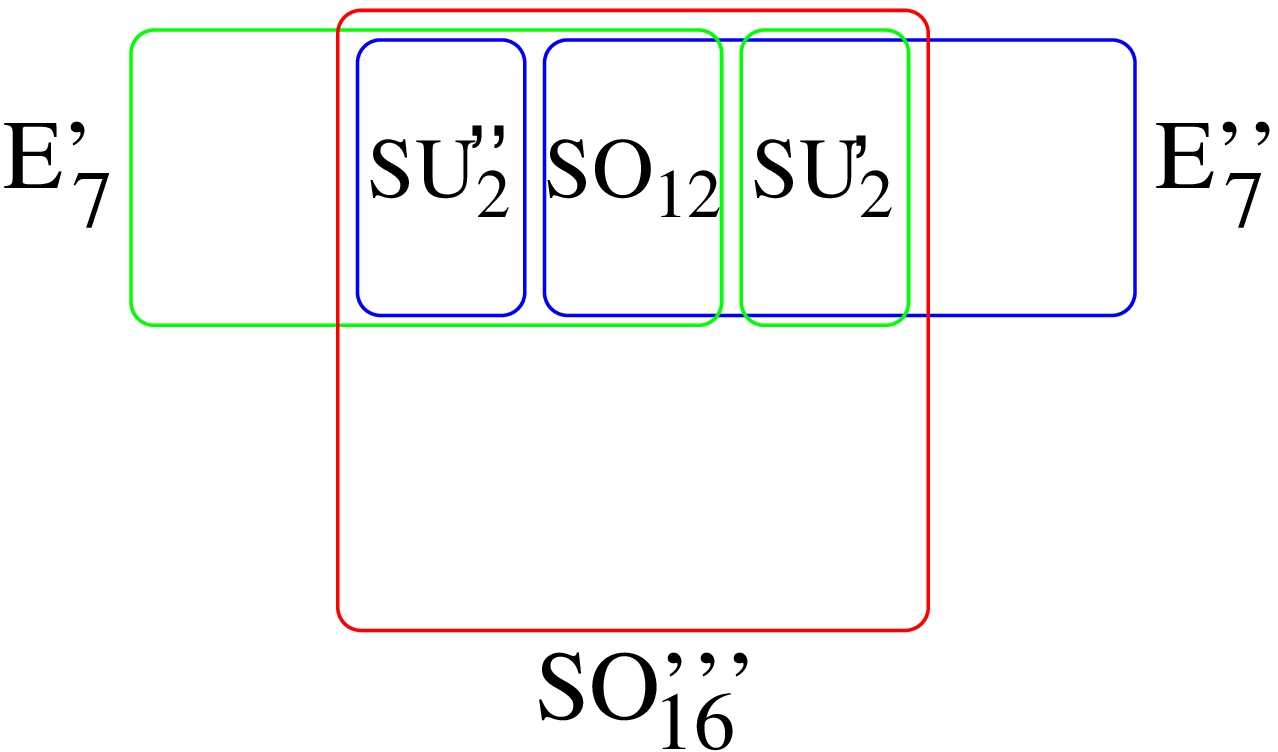}\\[.4in]
 \caption{Embedding diagram depicting the branching $E_8\to
 SO_{12}\times SU_2\times SU_2$.}
 \label{embed}
 \vspace{1in}
 \includegraphics[width=2.5in,angle=0]{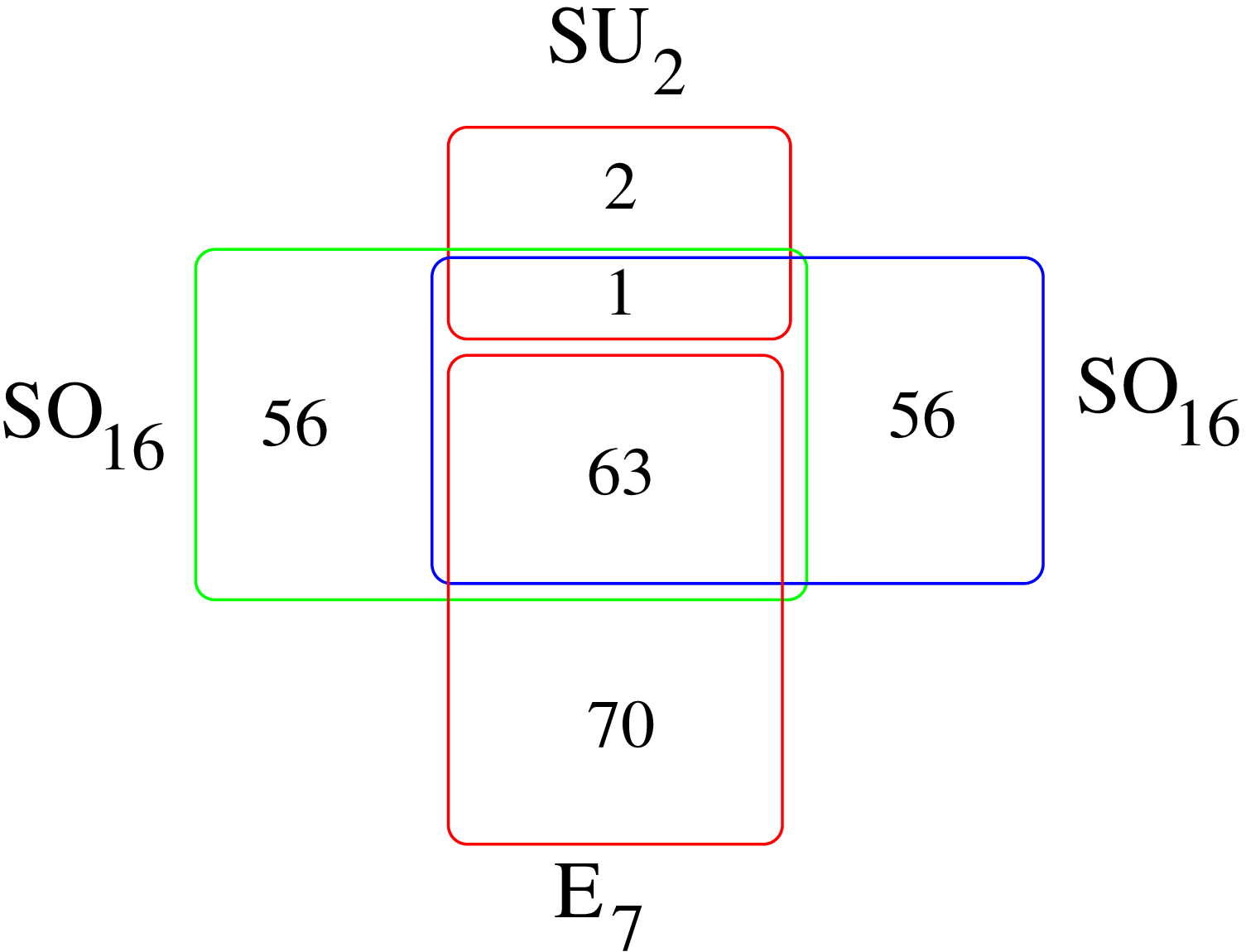} \hspace{.4in}
 \includegraphics[width=2.5in,angle=0]{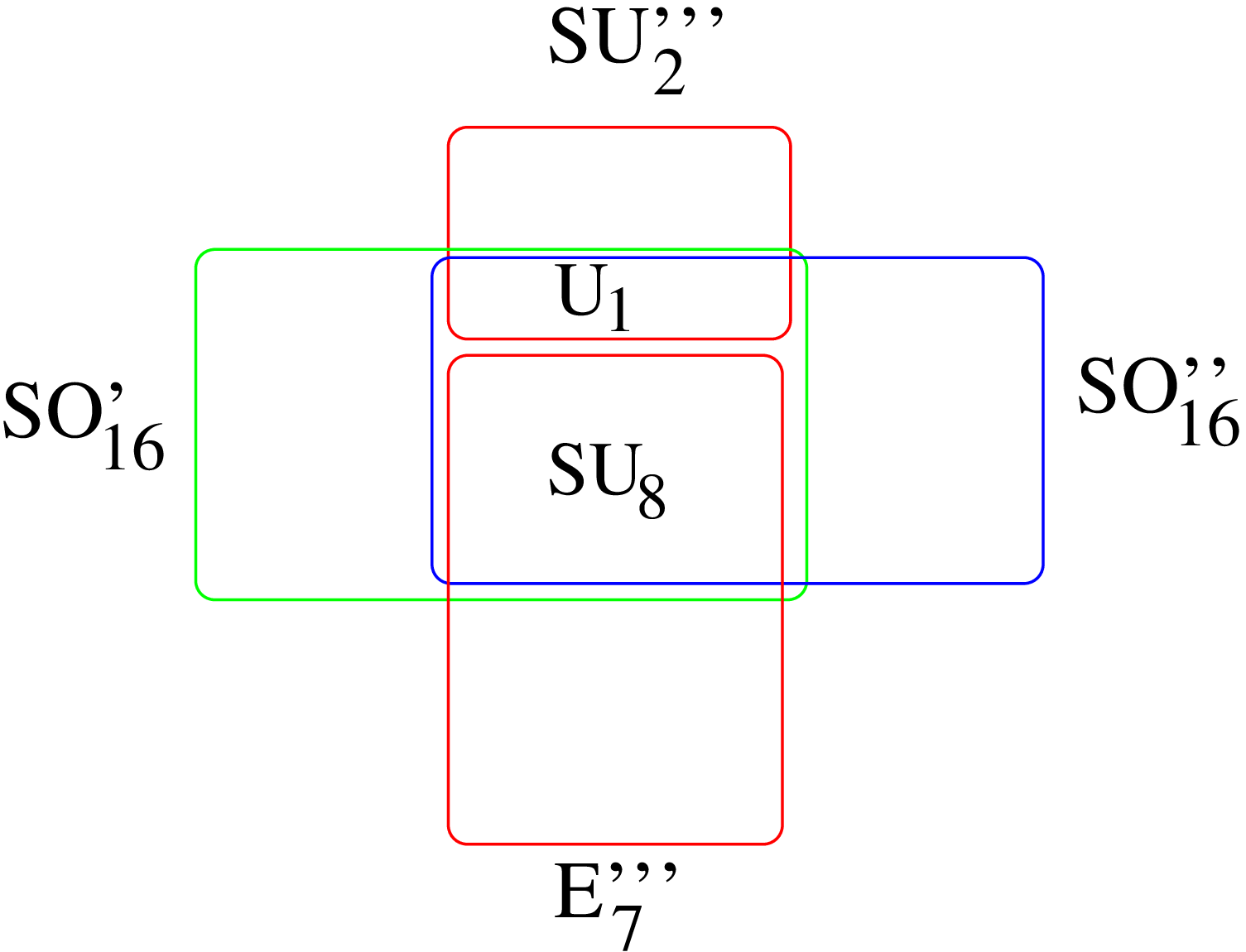}\\[.4in]
 \end{center}
 \caption{Embedding diagram depicting the branching $E_8\to
 SU_8\times U_1$.}
 \label{embed2}
 \end{figure}

In the ``upstairs" embedding,
the group ${\cal G}_\a=E_7\times SU_2$
is not the same subgroup of $E_8$ described by
${\cal G}_\g=E_7\times SU(2)$.  Similarly, in the ``downstairs"
embedding, the group ${\cal G}_\a=SO(16)$
is not the same subgroup of $E_8$ described by
${\cal G}_\g=SO(16)$.  The various embeddings are usefully
depicted by a specialized diagram, which we will call an ``embedding diagram".
These illustrate how the various groups ${\cal G}_\a$, ${\cal G}_\b$,
${\cal G}_\g$ and $H$ are embedded inside of $E_8$, and comprise
two-dimensional ``maps" of $E_8$ in which regions denoted by closed curves
correspond to specified subgroups.
For instance, the ``upstairs" case described by Table \ref{branch1}
has the embedding diagram shown in Figure \ref{embed},
which we have drawn in two incarnations, side by side, with different useful
aspects of data entered in each incarnation.

In Figure \ref{embed}, the subgroup
${\cal G}_\a$ is represented by the region surrounded by a green
boundary, the group ${\cal G}_\b$ is with a red boundary, and the
group ${\cal G}_\g$ with a blue boundary. The region with a
particularly colored boundary we call a bubble. Thus, the green
bubble includes the 133+3=136 generators of one of the $E_7\times
SU_2$ subgroups.
In the left-hand diagram the dimensionalities of
the subgroups are indicated and in the right-hand diagram the
identity of the subgroups are indicated.  Thus, the numbers in the
left-hand diagram count the $E_8$ generators. The fact that
$\a\b\g=1$ implies restrictions which can be interpreted directly
on these embedding diagrams. For instance, since each column in
Table \ref{branch1} must have entries whose product is plus one,
it follows that there can be either three plusses and no minuses
or one plus and two minuses in any column. There is no other
possibility. As a result, the embedding diagram will include
regions which are enclosed by all three bubbles or by only one.
Furthermore, every one of the 248
generators of $E_8$ are enclosed in at least one bubble.
These topological restrictions on the embedding diagram encapsulate
the consistency requirement imposed by $\a\b\g=1$.
The ``downstairs" branching, described by Table \ref{branch2},
is likewise described by the embedding diagram shown in
Figure \ref{embed2}.
The set
of generators enclosed in all three bubbles corresponds to the
group $H$.  Thus, we can read from figure \ref{embed} the complete
embedding of all four subgroups ${\cal G}_\a\subset E_8$,
${\cal G}_\b\subset E_8$, ${\cal G}_\g\subset E_8$ and $H\subset E_8$.

\section{The Four-Dimensional Spectrum}

>From the information included in the branching table and the
embedding diagram, it is straightforward to determine the
spectrum seen by the four-dimensional intersection which arises
from the ten-dimensional $E_8$ fields.  (We will call these the
$10\to 4$ fields because of useful comparisons to be made
later on.)
We decompose the ten-dimensional vector fields into four-dimensional
fields as
$A_{\bar{\mu}}^a=\{A_\mu^a\,,\,\Phi_1^a\,,\,\Phi_2^a\,,\,\Phi_3^a\,\}$,
where $A_\mu^a$ are four dimensional vectors and
$\Phi_i^a$ are three sets of complex scalars.
A given $\Phi_i^a$ combines with the
four-dimensional vector to form a six-dimensional
vector.   Thus, $\Phi_1^a$ describes a four-dimensional scalar,
but corresponds to vector degrees of freedom on the six-dimensional
$\a$-plane (but as scalars on the $\b$ and $\g$ six-planes).
Similarly, $\Phi_2^a$ is associated with the
$\b$-plane and  $\Phi_3^a$ is associated with the $\g$
plane.  The three generators
$\a$, $\b$ and $\g$ act on the tensor components (i.e. the lower index)
of $A_{\bar{\mu}}^a$ via multiplication by
signs as listed in Table \ref{veca}.
(This describes the tensorial transformation derived from the quotient
group action shown in Table \ref{modone}.)
 \begin{table}
 \begin{center}
 \begin{tabular}{c|ccc}
 & $\a$ & $\b$ & $\g$ \\[.1in]
 \hline
 $A_\mu^a$  & $+$ & $+$ & $+$ \\[.1in]
 $\Phi_1^a$ & $-$ & $-$ & $+$ \\[.1in]
 $\Phi_2^a$ & $+$ & $-$ & $-$ \\[.1in]
 $\Phi_3^a$ & $-$ & $+$ & $-$
 \end{tabular}\\[.2in]
 \end{center}
 \caption{The {\it tensorial} action of the three elements $\a$,
 $\b$ and $\g$ on the components of a ten-dimensional vector field.
 Note: we have not included the action on the group index $a$
 (i.e. the {\it lattice} action) in this table;
 these are described in Table \ref{branch1} or \ref{branch2}.}
 \label{veca}
 \end{table}

The full action of the quotient group $(\Z_2)^3$ on the
$E_8$-valued fields $A_{\bar{\mu}}^a$ includes not only the tensorial action
(the lower index), but also
the action on the $E_8$ root lattice (codified by the upper index).
The components of $A_{\bar{\mu}}^a$ partition into vector and scalar fields,
but also into different representations of $H$, as tabulated in the
rows of Table \ref{branch1} and Table \ref{branch2}.  Each partition
transforms according to the {\it product} of the relevant entry in
Table \ref{veca} with the corresponding entry in the appropriate branching table,
either Table \ref{branch1} or Table \ref{branch2} depending on whether
the four-plane is ``upstairs" of ``downstairs" respectively.
On the four-plane the surviving components are those
which transform trivially under {\it each}
generator $\a$, $\b$ and $\g$; those transforming non-trivially under
any one of these are projected out. Thus, surviving fields transform with
an overall plus sign under {\it each}
of the generating elements $\a$, $\b$ and $\g$, taking both the
tensorial action and the $E_8$ lattice action together.
Therefore, we resolve the $10\to 4$ spectrum by
comparing Table \ref{branch1} with Table \ref{veca}, and matching
the rows.  Surviving vectors
transform in representations of $H$ indicated by
$(+++)$, while surviving
complex scalars transform in representations
indicated by $(--+)$, $(+--)$ or $(-+-)$.  Surviving vectors
live in $N=1$ vector supermultiplets and surviving
scalars live in chiral multiplets.

The $10\to 4$ spectrum seen by an upstairs four-plane
(i.e. one whose branching is described by Table \ref{branch1} or
by Figure \ref{embed}) involves 66+3+3=72
vector multiplets transforming as the adjoint of $SO_{12}\times
SU_2\times SU_2$ and 64+48+64=176 chiral
multiplets transforming as
\brr  ({\bf 32}',{\bf 1},{\bf 2})\oplus
      ({\bf 12},{\bf 2},{\bf 2})\oplus
      ({\bf 32},{\bf 2},{\bf 1}) \,.
\err
The respective terms in this decomposition correspond to
6D scalars on the $\a$, $\b$ and $\g$ fixed-planes.  Note that
the respective multipicities can also be read off of the
embedding diagram from the bubbled regions outside of the
total intersection.

The $10\to 4$ spectrum seen by an downstairs four-plane
(i.e. one whose branching is described by Table \ref{branch2} or
by Figure \ref{embed2}) involves 63+1=64
vector multiplets transforming as the adjoint of $SU(8)\times U(1)$
and 56+56+70+2=184 chiral
multiplets transforming as
\brr  {\bf 28}_{-1}\oplus {\bf\overline{28}}_{+1}\oplus
      {\bf 28}_{1}\oplus {\bf\overline{28}}_{-1}\oplus
      {\bf 70}_0\oplus {\bf 1}_{+2}\oplus {\bf 1}_{-2} \,.
\err
Note again that
the respective multipicities can also be read off of the
embedding diagram from the bubbled regions outside of the
total intersection.

\section{Remaining Twisted Sectors}
There are two more sources of twisted matter for the orbifold which we
are discusing.  The first are six-dimensional fields
added to some of the six-planes in order to
cancel purely six-dimensional anomalies.  The second are seven-dimensional
fields added to the seven-planes, because these too are
needed to cancel the six-dimensional anomalies.  This last statement is a
subtle one, which was described in \cite{mlo, phase, hetk3}
and summarized in \cite{chern}.  In this section we discuss
all of the remaining twisted states in this orbifold.

\subsection{Six-Dimensional Fields}
As described in \cite{phase, hetk3} the six dimensional anomaly is cancelled
without six-dimensional twisted fields
for the case of $E_8\to E_7\times SU(2)$ breakdown.  For $E_8\to SO(16)$ breakdown,
however, it is necessary to add six-dimensional hypermultiplets.
These transform as $\ft12({\bf 16},{\bf 2})$ under $SO(16)\times SU(2)$
where the $SU(2)$ factor is associated with the adjacent seven-plane.
(The extra $SU(2)$ is a subgroup of the ``other" $E_8$ factor,
as we recall in the next subsection.)  The six-dimensional twisted
fields reduce to chiral multiplets in four dimensions.  We refer to these
fields as the $6\to 4$ spectrum

For the ``upstairs" four planes, only the $\b$-invariant six-planes
support the branching $E_8\to SO(16)$.  Because of this, we add hypermultiplets
transforming as $\ft12({\bf 16},{\bf 2})$ to the upstairs $\b$-planes.
The four-plane intersections see these fields as $N=1$
chiral multiplets
\footnote{The $\ft12$ on the hypermultiplet representation serves to reduce
the four scalars in the hypermultiplet into the two real
scalars which combine to the one complex scalar in the
four-dimensional chiral multiplet.}.
The representation branches to $H\times SU(2)$
according to
\brr SO_{16}\times SU_2 &\to& SO_{12}\times SU_2\times SU_2\times SU_2
     \nonumber\\
     ({\bf 16},{\bf 2}) &\to&
     ({\bf 12},{\bf 1},{\bf 1},{\bf 2})\oplus
     ({\bf 1},{\bf 2},{\bf 2},{\bf 2}) \,;
\label{clo}\err
the ``upstairs" $6\to 4$
spectrum includes chiral multiplets transforming under
$SO(12)\times SU(2)\times SU(2)$
according to the right hand side of (\ref{clo}).

For the ``downstairs" four planes, the $\a$-planes and the $\g$-planes
support $E_8\to SO(16)$ branchings.  Because of this, we add hypermultiplets
to the downstairs $\a$-planes and the downstairs $\g$-planes.
In either case, the associated twisted matter
branches to the four-dimensional gauge group $H\times SU(2)$
according to
\brr SO_{16}\times SU_2 &\to& SU(8)\times U(1)\times SU(2)
     \nonumber\\
     ({\bf 16},{\bf 2}) &\to&
     ({\bf 8},{\bf 2})_0\oplus
     ({\bf\overline{8}},{\bf 2})_0 \,;
\label{clo2}\err
the ``downstairs" $6\to 4$
spectrum includes chiral multiplets transforming under
$SU(8)\times SU(2)\times U(1)$
according to the right hand side of (\ref{clo}).
(A more precise accounting of {\it which} $SU(2)$ factors are being referred to
in each case is tabulated below in Table \ref{specd},
in a manner which will be explained.)

\subsection{Seven-Dimensional Fields}
The seven-planes corresponding to the group
elements $\a\b$, $\b\g$ and $\a\g$ each carry vector multiplets.
For the $(\Z_2)^3$ orbifold, these each support $SU(2)$ vector multiplets.
These are chirally projected on the intersecting six-planes
corresponding to $\a$, $\b$ and $\g$ onto six-dimensional hyper or
vector multiplets in a way dictated by the cancellation of
six-dimensional anomalies.  The choice of projection onto the
vectors or hypers is resolved in \cite{phase, hetk3} for
each global abelian $S^1/\Z_2\times K3$ orbifold. Since the cases
under discusion include these simpler cases as sub-orbifolds, this
issue is already resolved.  The appropriate choices are indicated
by arrows in diagrams such as Figure \ref{octopus},
with ``$V$" or
``$H$" labels describing the appropriate local projection. To
determine the  four-dimensinal spectrum from these
fields, we have to further project from six-dimensions down to
four. The result is that the seven-dimensional vectors undergo a
projection $7\to 6\to 4$ of one of three sorts, $V\to H\to
H$, $V\to V\to V$ or $V\to V\to H$.  Further details are explained
in \cite{phase, hetk3}.  This determines the so-called $7\to 4$ fields.   Rather
than list these here, we include these in the all-encompassing tables
presented in the next section which summarize the various sectors
of the effective four-dimensional spectrum.

\section{Synopsis}
\begin{figure}
\begin{center}
\vspace{1in}
\includegraphics[width=7in,angle=0]{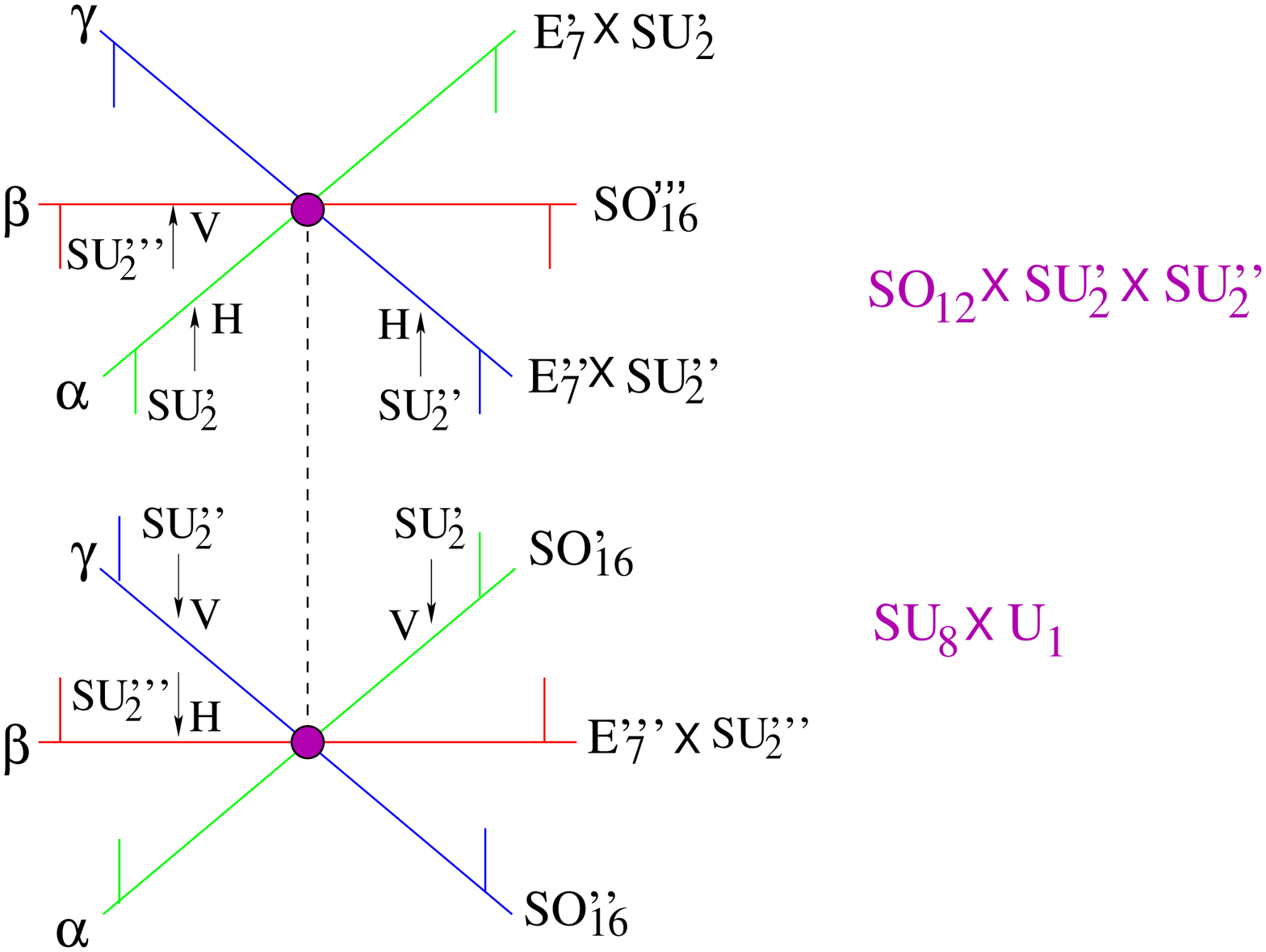}\\[1in]
\end{center}
\caption{Synopsis}
\label{octopus}
\end{figure}

By including $E_8$ Yang-Mills multiplets on the fixed ten-planes we
cancel the ten dimensional anomalies.  By properly accounting
for the action of the $(\Z_2)^3$ quotient group on the $E_8$ root lattice
we can describe
a consistent breakdown of these $E_8$ factors to appropriate
subgroups on the various fixed six-planes.  These are generically
either $E_7\times SU(2)$ subgroups or $SO(16)$ subgroups.  For the cases
where the six-dimensional gauge group is $SO(16)$ we must include
additional hypermultiplets on these six-planes.  In order to cancel the
six-dimensional anomalies we also include {\it seven}-dimensional
$SU(2)$ Yang-Mills multiplets on the fixed seven-planes.  At the
same time we assign magnetic $G$ charges to the six-planes to enable the
appropriate inflow anomalies.  (This last part of the story has been
largely suppressed in this paper because the relevant discussion
found in \cite{phase, hetk3} is unchanged in this context, save for
one global result related to this issue: that the upstairs and downstairs
breaking need be complementary in sense described above.)
The state of affairs is
largely summarized by Figure \ref{octopus}.

Figure \ref{octopus} summarizes the juxtaposition of all of the various
gauge group factors and the associated projections.  This figure is
a streamlined version of the waterwheel diagram shown in
Figure \ref{waterwheel}, with a few lines removed and with extra
data drawn in.  The two purple dots in this diagram depict one
``upstairs" four-plane and one ``downstairs" four plane, as well as lines
representing each of the six-planes which intersect at these points.
The groups ${\cal G}_\a$, ${\cal G}_\b$ and ${\cal G}_\g$ are indicated
to the immediate right of the relevant six-planes, and the group $H$ is indicated
to the far right for each of the cases, upstairs and downstairs.
The orbifold consists of an aggregation of 32
regions such as the one shown in this figure.

Having reconciled all of the twisted states, be they ten, six or seven
dimensional, into the four dimensional projections seen by a particular
four plane, the one remaining issue is to study an additional four-dimensional
anomaly which might arise due to these fields.  In four dimensions, the only
type of anomaly is a gauge anomaly.  Generally, such will arise
due to chiral coupling of the ten, six, or seven dimensional
fields to the four dimensional gauge currents.  If there is a four-dimensional
gauge anomaly, this must be resolvable by the addition of purely four-dimensional
twisted fields.

In order to determine the remaining anomaly we can use index theorems.
However, this requires special care, since the index theorem results usually employed
for anomaly calculations decribe anomalies due to the chiral coupling of fields with
certain dimensionality to currents of that same dimensionality.  As
emphasized in \cite{hw2} and used extensively in \cite{mlo,phase, hetk3, chern},
this is not necessarily the state of affairs in orbifolds, where the index theory
results must be modified by certain divisors which are corellated with the
multiplicities of fixed-planes.  We have summarized the effective
spectrum of twisted fields as seen by particular four-dimensional
intersections, for the $(\Z_2)^3$ orbifold,
in Tables \ref{specu} and \ref{specd}, where we have
also included the relevant multiplicity divisor for the indicated representations.
The fractions which appear in these
tables, therefore, indicate the number of four-planes over which a given higher
dimensional twisted state is distributed.  In order to compute the
effective low energy theory obtained by letting all 64 four-planes
coelesce, these fractions usefully account for the appropriate
multiplicities in the spectrum.

\begin{table}
\begin{center}
\vspace{2in}
\begin{tabular}{|c||c|c|c|}
\hline
&&&\\[-.1in]
&10D & 6D & 7D \\[.1in]
\hline
Chiral &$\ft{1}{64}({\bf 32},{\bf 2},{\bf 1},{\bf 1})$ &
 $\ft14({\bf 12},{\bf 1},{\bf 1},{\bf 2})$  &
 $\ft18({\bf 1},{\bf 3},{\bf 1},{\bf 1})$ \\[.1in]
&$\ft{1}{64}({\bf 12},{\bf 2},{\bf 2},{\bf 1})$  &
 $\ft14({\bf 1},{\bf 2},{\bf 2},{\bf 2})$  &
 $\ft18({\bf 1},{\bf 1},{\bf 3},{\bf 1})$ \\[.1in]
&$\ft{1}{64}({\bf 32'},{\bf 1},{\bf 2},{\bf 1})$ & & \\[.1in]
\hline
Vector &$\ft{1}{64}({\bf 66},{\bf 1},{\bf 1},{\bf 1})$ & &
 $\ft18({\bf 1},{\bf 1},{\bf 1},{\bf 3})$ \\[.1in]
&$\ft{1}{64}({\bf 1},{\bf 3},{\bf 1},{\bf 1})$ & & \\[.1in]
&$\ft{1}{64}({\bf 1},{\bf 1},{\bf 3},{\bf 1})$ & & \\[.1in]
\hline
\end{tabular}\\[.2in]
\caption{The untwisted spectrum, in terms of four-dimensional
chiral multiplets and vector multiplets at one of the ``upstairs"
orbifold four-planes expressed in terms of representations of
$SO_{12}\times SU_2'\times SU_2''\times SU_2'''$.}
\label{specu} \vspace{3in}
\end{center}
\end{table}

\begin{table}
\begin{center}
\begin{tabular}{|c||c|c|c|}
\hline
&&&\\[-.1in]
&10D & 6D & 7D \\[.1in]
\hline
Chiral &$\ft{1}{64}({\bf 28},{\bf 1},{\bf 1})_{+1}$ &
 $\ft14({\bf 8},{\bf 2},{\bf 1})_0$  &
 $\ft18({\bf 1},{\bf 1},{\bf 1})_0$ \\[.1in]
&$\ft{1}{64}({\bf\overline{28}},{\bf 1},{\bf 1})_{-1}$  &
 $\ft14({\bf\overline{8}},{\bf 2},{\bf 1})_0$  &
 $\ft18({\bf 1},{\bf 1},{\bf 1})_{+2}$ \\[.1in]
&$\ft{1}{64}({\bf 70},{\bf 1},{\bf 1})_0$ &
 $\ft14({\bf 8},{\bf 1},{\bf 2})_0$ &
 $\ft18({\bf 1},{\bf 1},{\bf 1})_{-2}$ \\[.1in]
 &$\ft{1}{64}({\bf 1},{\bf 1},{\bf 1})_{-2}$ &
  $\ft14({\bf \overline{8}},{\bf 1},{\bf 2})_0$ &  \\[.1in]
&$\ft{1}{64}({\bf 1},{\bf 1},{\bf 1})_{+2}$ &  &  \\[.1in]
&$\ft{1}{64}({\bf 28},{\bf 1},{\bf 1})_{-1}$ &  &  \\[.1in]
&$\ft{1}{64}({\bf \overline{28}},{\bf 1},{\bf 1})_{+1}$ &  &  \\[.1in]
\hline
Vector & $\ft{1}{64}({\bf 63},{\bf 1},{\bf 1})_0$ & &
$\ft18({\bf 3},{\bf 1},{\bf 1})_0$ \\[.1in]
&$\ft{1}{64}({\bf 1},{\bf 1},{\bf 1})_0$ & &
$\ft18({\bf 1},{\bf 3},{\bf 1})_0$ \\[.1in]
\hline
\end{tabular}\\[.2in]
\caption{The untwisted spectrum, in terms of four-dimensional
chiral multiplets and vector multiplets at one of the
``downstairs" orbifold four-planes expressed in terms of
representations of $SU_8\times SU_2'\times SU_2''\times U_1$.}
\label{specd}
\end{center}
\end{table}

\section{The Four-Dimensional Anomaly}

It remains to study the gauge anomaly seen locally at the
four dimensional intersection.  In typical orbifolds, there will be a
localized four-dimensional gauge anomaly.  Further four-dimensional
twisted states would need
be added to cancel this. But in the relatively simple
$(\Z_2)^3$ orbifold described in this paper, there is no
four-dimensional anomaly which needs to be cured in this way.
This is because a four dimensional gauge anomaly is only
generated by chiral fields transforming in complex representations.
Otherwise the third index of the representation vanishes, so that
${\rm tr}\,F^3=0$.  Since gauge anomalies in four dimensions are
proportional to precisely this trace, there is no gauge anomaly
induced by fields transforming in real representations.  Since all of the
representations which appear in Tables \ref{specu} and \ref{specd}
are real, we have completed the anomaly cancellation program by adding in
the ten, six and seven dimensional fields sumarized in these tables.

In the limit that the compact dimensions become very small, the effective
four-dimensional spectrum is obtained by adding up the contributions from
all 64 fixed four-planes.  For the $(\Z_2)^3$ orbifold the results are sumarized
in Table \ref{result}.

\begin{table}
\begin{center}
\begin{tabular}{|c|c|}
\hline
& \\[-.1in]
$({\bf 32},{\bf 1},{\bf 2},{\bf 1})_0$ &
$16\,({\bf 12},{\bf 1},{\bf 1},{\bf 1})_{+1}$
\\[.1in]
$({\bf 12},{\bf 1},{\bf 2},{\bf 2})_0$ &
$16\,({\bf 12},{\bf 1},{\bf 1},{\bf 1})_{-1}$
\\[.1in]
$({\bf 32'},{\bf 1},{\bf 1},{\bf 2})_0$ &
$16\,({\bf 1},{\bf 1},{\bf 2},{\bf 2})_{+1}$
\\[.1in]
& $16\,({\bf 1},{\bf 1},{\bf 2},{\bf 2})_{-1}$  \\[.1in]
\hline
$({\bf 1},{\bf 28},{\bf 1},{\bf 1})_{+1}$ &
$16\,({\bf 1},{\bf 8},{\bf 2},{\bf 1})_{0}$ \\[.1in]
$({\bf 1},{\bf\overline{28}},{\bf 1},{\bf 1})_{-1}$&
$16\,({\bf 1},{\bf\overline{8}},{\bf 2},{\bf 1})_{0}$
\\[.1in]
$({\bf 1},{\bf 70},{\bf 1},{\bf 1})_{0}$&
$16\,({\bf 1},{\bf 8},{\bf 1},{\bf 2})_{0}$
\\[.1in]
$({\bf 1},{\bf 1},{\bf 1},{\bf 1})_{-2}$ &
$16\,({\bf 1},{\bf\overline{8}},{\bf 1},{\bf 2})_{0}$
\\[.1in]
$({\bf 1},{\bf 1},{\bf 1},{\bf 1})_{+2}$ &  \\[.1in]
$({\bf 1},{\bf 28},{\bf 1},{\bf 1})_{-1}$ &  \\[.1in]
$({\bf 1},{\bf\overline{28}},{\bf 1},{\bf 1})_{+1}$ &  \\[.1in]
\hline
\end{tabular}\\[.2in]
\caption{The representation content of D=4, N=1 chiral multiplets
in Model 1, in terms of the gauge group $SO_{12}\times SU_8\times
SU_2\times SU_2\times U_1$.  The fields above the bar arise from
the upstairs sector. The fields below the bar arise from the
downstairs sector.  Fields with a multiplicity of 16 arise as
twisted 6D fields.  Note that all 7D fields cancel in the limit
that all fixed-planes coalesce.}
\label{result}
\end{center}
\end{table}

\section{Conclusions}

In this paper, we have shown explicitly how a four dimensional $N=1$
Yang-Mills theory can be determined from anomaly matching on
the fixed-planes of an {\it M}-theory orbifold.
One interesting aspect of this analysis is the manner in which
constraints from gravitational anomalies impinge on the
four dimensional spectrum, despite the fact that there are no
gravitational anomalies in purely four
dimensional field theories.

In our on-going work we are developing a systematic scan of all possible
orbifolds obtained as quotients $(\R^7/\Lambda)/G$ for each possible
lattice $\Lambda$ and for each possible choice of
quotient group $G\subset {\rm Aut}(\Lambda)$.  For a given orbifold
constructed in this way we select those which have supercharges
preserved on the fixed-planes, and then ascertain the fixed-plane
twisted spectra needed to cancel all local anomalies.  The more
interesting cases involve non-abelian quotients.  The orbifold described
in this paper is the abelian orbifold of smallest possible group order
yielding $N=1$ SUSY in four dimensions.  One purpose of this
paper has been to present a context for some of the rudiments of the larger algorithm
which we are implementing on larger class of orbifolds, in the hopes of
finding a phenomenologically compelling effective field theory limit
of {\it M}-theory.

\vspace{.5in}
{\Large {\bf Acknowledgements}}\\[.1in]
M.F. would like to thank Dieter L{\" u}st for warm hospitality at
Humboldt University where a portion of this manuscript was prepared,
and also the organizers of the workshop {\it Dualities: A Math/Physics
Collaboration}, ITP, Santa Barbara.

\end{document}